\documentclass[conference,a4paper]{IEEEtran}

\IEEEoverridecommandlockouts
\usepackage{cite}
\usepackage{amsmath,amssymb,amsfonts}
\usepackage{graphicx}
\usepackage{textcomp}
\usepackage{xcolor}
\usepackage{algorithm}
\usepackage{algorithmicx}
\usepackage{algpseudocode}
\usepackage{subcaption}
\usepackage{booktabs}
\usepackage[final]{microtype}
\usepackage{setspace}
\usepackage{enumitem}
\usepackage{tikz}
\usepackage{times}
\usepackage{newtxtext,newtxmath}

\setlist[itemize]{noitemsep, topsep=0pt, leftmargin=*}
 \usepackage[a4paper, total={184mm,260mm}]{geometry}

\def\BibTeX{{\rm B\kern-.05em{\sc i\kern-.025em b}\kern-.08em
    T\kern-.1667em\lower.7ex\hbox{E}\kern-.125emX}}
\begin{document}

% Command for circled numbers
\newcommand*\circled[1]{\tikz[baseline=(char.base)]{
            \node[shape=circle,draw,inner sep=1pt,fill=black,text=white] (char) {#1};}}

\title{RapidOMS: FPGA-based Open Modification Spectral Library Searching with HD Computing \vspace{-0.4cm}}

% \author{
%     \IEEEauthorblockN{
%         Sumukh Pinge\IEEEauthorrefmark{1},
%         Weihong Xu\IEEEauthorrefmark{1},
%         Wout Bittremieux\IEEEauthorrefmark{2},
%         Niema Moshiri\IEEEauthorrefmark{1},
%         Sang-Woo Jun\IEEEauthorrefmark{3},
%         Tajana Rosing\IEEEauthorrefmark{1}
%     }
%     \IEEEauthorblockA{\IEEEauthorrefmark{1}University of California San Diego, La Jolla, CA 92093, USA\\
%     \{spinge, wexu, a1moshir, tajana\}@ucsd.edu}
%     \IEEEauthorblockA{\IEEEauthorrefmark{2}University of Antwerp, 2000 Antwerpen, Belgium\\
%     wout.bittremieux@uantwerpen.be}
%     \IEEEauthorblockA{\IEEEauthorrefmark{3}University of California, Irvine, Irvine, CA 92697, USA\\
%     swjun@ics.uci.edu}
% }

\author{
    Sumukh Pinge\IEEEauthorrefmark{1},
    Weihong Xu\IEEEauthorrefmark{1},
    Wout Bittremieux\IEEEauthorrefmark{2},
    Niema Moshiri\IEEEauthorrefmark{1},
    Sang-Woo Jun\IEEEauthorrefmark{3},
    Tajana Rosing\IEEEauthorrefmark{1}
    \\
    \IEEEauthorrefmark{1}University of California San Diego, La Jolla, CA 92093, USA\\
    \{spinge, wexu, a1moshir, tajana\}@ucsd.edu\\
    \IEEEauthorrefmark{2}University of Antwerp, 2000 Antwerpen, Belgium, wout.bittremieux@uantwerpen.be\\
    \IEEEauthorrefmark{3}University of California, Irvine, Irvine, CA 92697, USA, swjun@ics.uci.edu
}

\maketitle
\begin{abstract}
Mass spectrometry (MS) is essential for protein analysis but faces significant challenges with large datasets and complex post-translational modifications, resulting in difficulties in spectral identification. Open Modification Search (OMS) improves the analysis of these modifications. We present RapidOMS, a solution leveraging the Samsung SmartSSD, which integrates SSD and FPGA in a near-storage configuration to minimize data movement and enhance the efficiency of large-scale database searching. RapidOMS employs hyperdimensional computing (HDC), a brain-inspired, high-dimensional data processing approach, exploiting the parallel processing and low-latency capabilities of FPGAs, making it well-suited for MS. Utilizing the parallelism and efficiency of bitwise operations in HDC, RapidOMS delivers up to a 60x speedup over the state-of-the-art (SOTA) CPU tool ANN-Solo and is 2.72x faster than the GPU tool HyperOMS. Furthermore, RapidOMS achieves an 11x improvement in energy efficiency compared to conventional systems, providing scalable, energy-efficient solutions for large-scale proteomics applications and advancing the efficient processing of proteomic data.
\end{abstract}
 

\section{Introduction}

Mass spectrometry (MS) holds a central role in proteomics research, particularly in analyzing the intricate structures of proteins in various biological samples, providing detailed insights crucial for advancing our understanding of biological processes and disease mechanisms \cite{Angel2012, Kim2014, Chick2015, Boisvert2012}. 

\subsection{Challenges of Large-scale Spectral Library Searching}
\begin{figure}[t]
    \centering
    \begin{subfigure}{.6\columnwidth}
        \centering
        % Adjust the height parameter as needed
        \includegraphics[width=\linewidth]{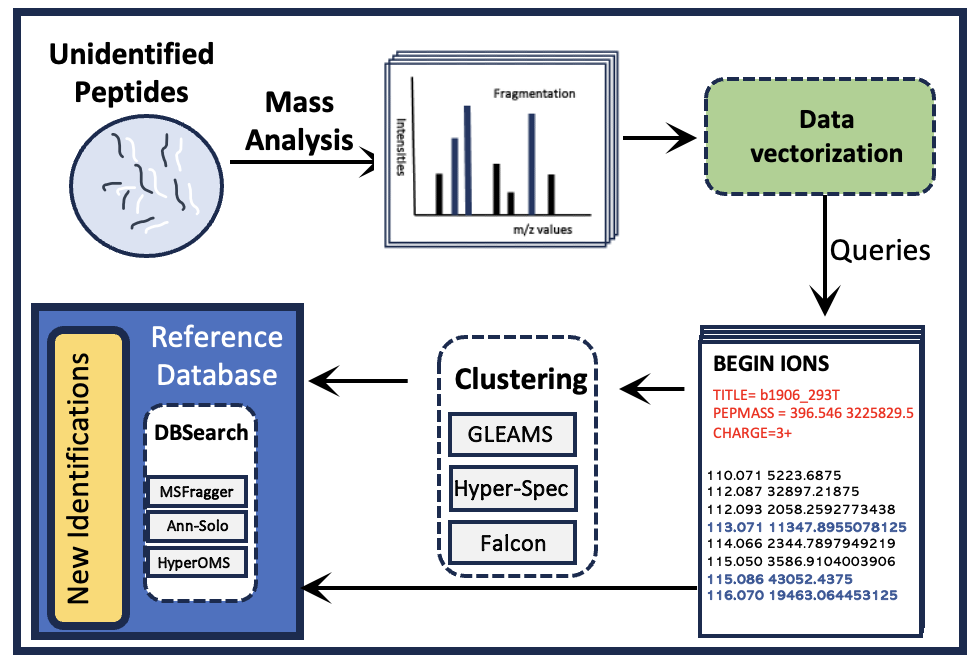} 
        \caption{Overall Flow}
        \label{fig:pipeline}
    \end{subfigure}%
    \hspace{0.04\columnwidth}
    \begin{subfigure}{.335\columnwidth}
        \centering
        % Adjust the height parameter as needed
        \includegraphics[width=\linewidth]{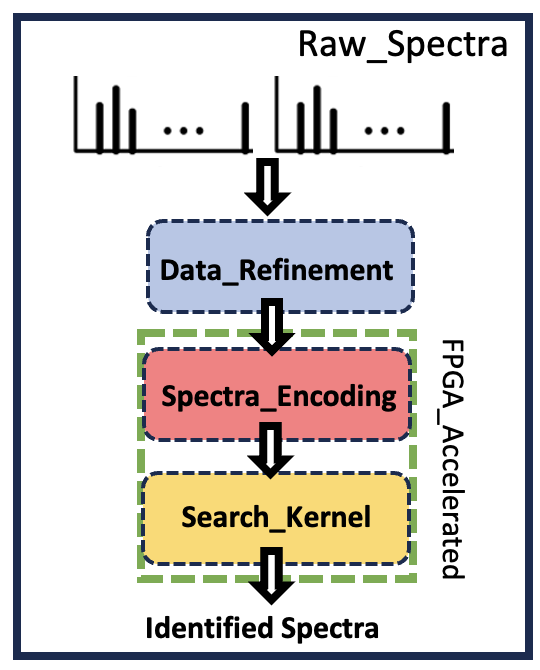}
        \caption{RapidOMS Flow}
        \label{fig:RapidOMSflow}
    \end{subfigure}
    \caption{MS Data Analysis Pipeline}
    \label{fig:mspipe}
\end{figure}

\noindent\textbf{Challenge 1-- Explosive Data Growth:} Rapid advancements in MS techniques have led to an exponential increase in data volume, with repositories like MassIVE accumulating over 588TB and PRIDE seeing a 10x increase in submissions over the past decade\cite{martens2005pride}. This enables comprehensive studies while also presenting significant challenges, particularly doing an efficient library search for new samples, essential for drug discovery and biomarker identification in personalized medicine\cite{Duarte2016}. The MS flow (Fig. \ref{fig:pipeline}) begins with ionization, where molecules in a sample are charged and sorted by their mass-to-charge (m/z) ratios in a mass analyzer. The generated spectra, crucial for understanding molecular composition, are converted into digital representations, refined, and clustered. This refined data is then matched against spectral libraries using various MS search tools\cite{annsolo,hyperoms,Kang2023,Kong2017}.  This is a compute-intensive phase in protein identification.  In metaproteomics, the study of proteins in microbiomes, the need for extensive library search is amplified, managing complex protein interactions across diverse environments and scaling datasets to gigabytes or terabytes\cite{mistle}.

\noindent\textbf{Challenge 2-- PTM-Induced Complexity:}
Post-translational modifications (PTMs) add to the complexity of the database search process\cite{unmatched}. Open Modification Search (OMS)\cite{annsolo} was developed to manage these challenges by accommodating vast  spectral variations, enhancing PTM analysis. Standard search uses a narrow precursor m/z tolerance, while OMS employs a wide precursor m/z tolerance, allowing it to select reference candidates even with mass shifts induced by modifications. However, OMS significantly increases computational demands by expanding the search space, affecting analysis sensitivity and duration. CPU-based solutions such as MSFragger\cite{Kong2017} use fragment ion indexing, resulting in lengthy run time. The CPU version of ANN-SoLo \cite{annsolo}, which employs nearest-neighbor search and shifted cosine similarity, also has complex execution pipelines and data parallelism limitations. HD computing-based GPU tools like HyperOMS\cite{hyperoms} enhance OMS efficiency but have limitations, such as performing exhaustive calculations for all references and queries before spectral identification, and facing scalability challenges due to limited memory capacity.

 \begin{figure}[t]
\centering
\includegraphics[width=\linewidth]{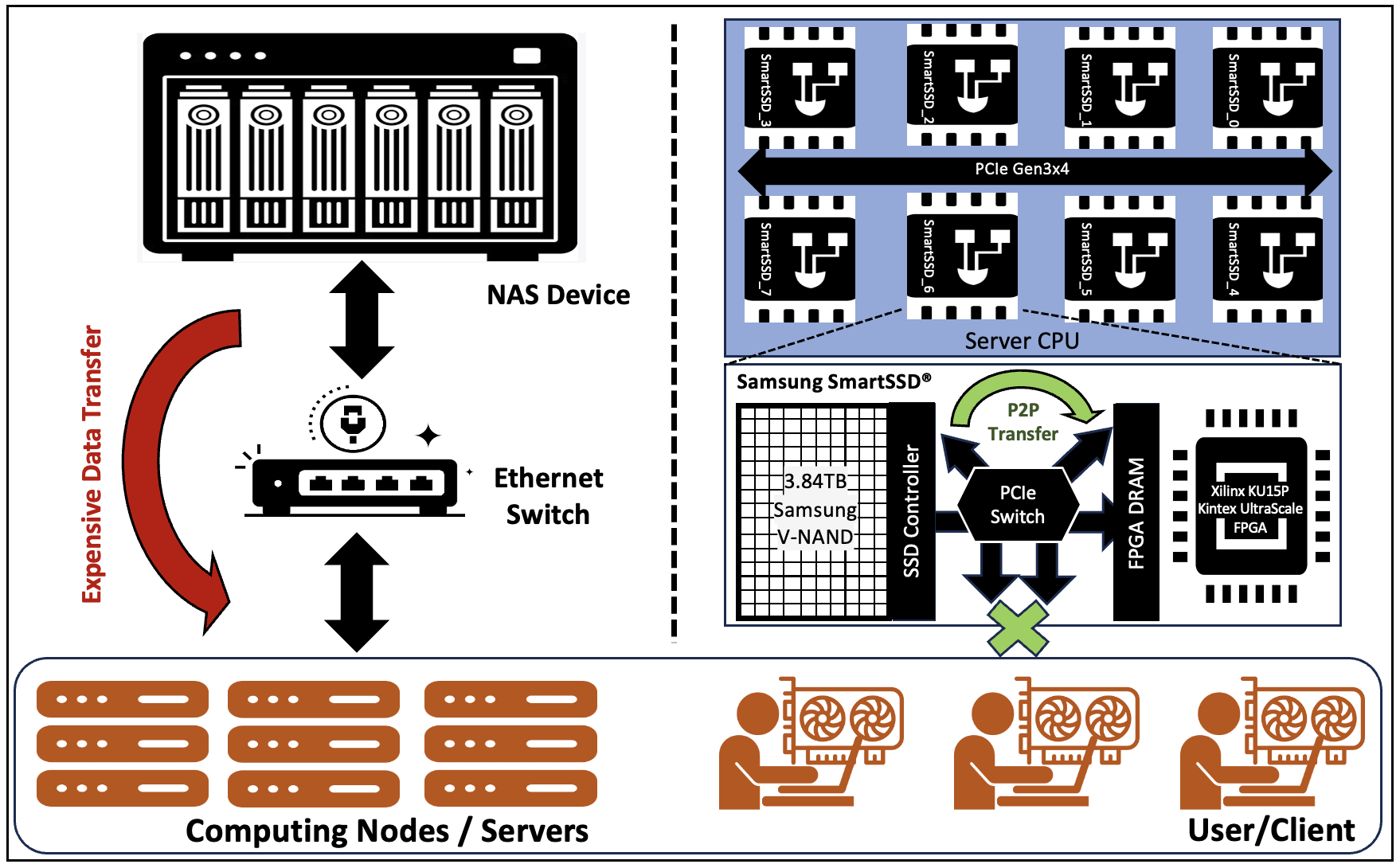}
\caption{SmartSSDs eliminates data transfers vs NAS Systems}
\label{fig:nas_diagram}
\end{figure}

\subsection{Motivations and Contributions of RapidOMS}
As datasets grow, research communities are turning towards Network Attached Storage (NAS) systems\cite{Meckel2014, MSUMassSpectrometry,RockefellerProteomics} to address storage requirements and data redundancy. However, this pivot introduces substantial inefficiencies in data processing workflows, as data must traverse the network to the host system before proceeding to the GPU’s DRAM, incurring significant latency and higher energy expenses (Fig. \ref{fig:nas_diagram})\cite{google,elnahas2020robust,Hoisie2013}. Near-storage (NS) solutions, particularly FPGA-based systems, offer scalability and reduced data handling overhead. Tools like MSAS\cite{msas} for preprocessing and SpecHD\cite{pinge2023spechd} for clustering have demonstrated significant performance improvements with near-storage compute, showing increased performance of up to 187× vs. CPU and 54× vs. GPU benchmarks. 

To address OMS-specific challenges, we introduce RapidOMS (fig.\ref{fig:RapidOMSflow}), an innovative solution utilizing compute-near-SSD integration and FPGA-accelerated techniques, significantly improving OMS performance. It leverages the Samsung SmartSSD®, which integrates SSD storage with FPGA computational power, minimizing data movement and enhancing efficiency. This approach tackles the issues of large data volumes and complex PTM analysis, marking advancement in MS data processing. Key contributions of RapidOMS include : 
\begin{enumerate}   
    \item RapidOMS leverages FPGA-accelerated SmartSSDs, achieving up to 60x speedup over the SOTA CPU tool ANN-Solo\cite{annsolo}, and 2.72x faster performance than SOTA GPU tool HyperOMS\cite{hyperoms} with compute-NS integration.
    \item Novel data caching and block-level optimizations between SmartSSD's NVMe SSD and DRAM cut down comparisons, boosting kernel runtime by 5.5x over HyperOMS\cite{hyperoms} in the time-sensitive domain of personalized healthcare.
    \item It outperforms GPU-based SOTA HyperOMS\cite{hyperoms} by 11x in energy efficiency, offering scalable, energy-efficient enterprise solutions for large-scale proteomics applications.
\end{enumerate}

\section{Proposed RapidOMS Acceleration Framework}

\subsection{Spectra Preprocessing and HD Encoding}

\noindent\textbf{Preprocessing: } In MS data analysis, refining raw spectral data and converting it into numerical vectors is essential, aligning with HDC-MS methodologies \cite{pinge2023spechd, hyperoms}.  The process begins by filtering out peaks with intensities below 1\% of the highest peak to isolate significant peaks from background noise. These peaks are vectorized by categorizing their m/z ratios into discrete bins, combining intensities within the same bin, crucial for capturing the spectrum’s characteristics. 

\begin{figure}[t]
\centering
\includegraphics[width=\linewidth]{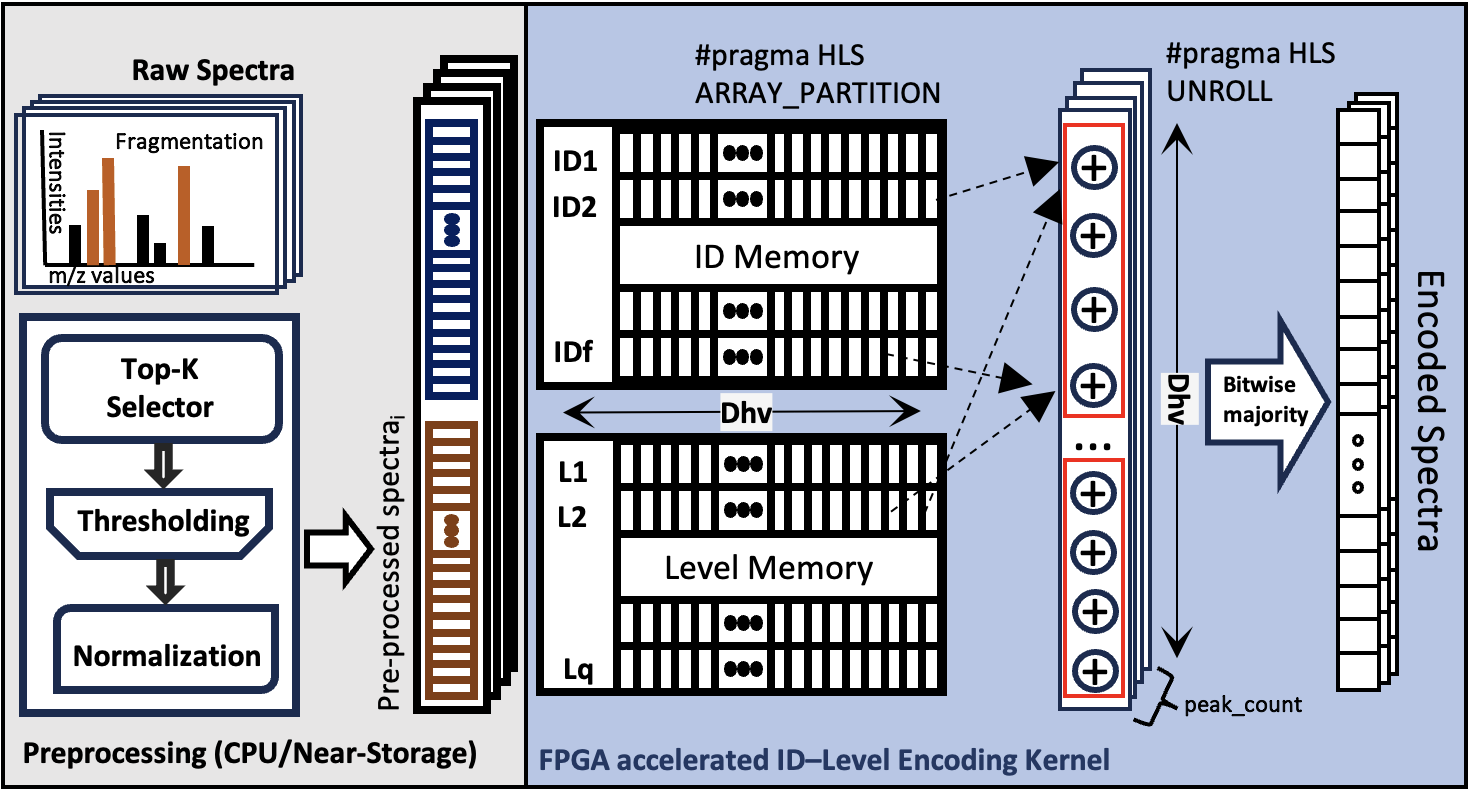}
\caption{MS Preprocessing + HLS-Optimized Spectra Encoder}
\label{fig:encoding}
\end{figure}

\noindent\textbf{HD Encoding: }Hyperdimensional computing (HDC)\cite{Kanerva2009} mimics the brain's information processing by operating in high-dimensional spaces with hypervectors (\textbf{HVs}). It utilizes element-wise operations like bundling and binding to generate dense, error-tolerant representations, ideal for parallel processing of proteomic data. To generate HD vectors, an HLS-optimized ID-Level encoding\cite{imani2017voicehd} kernel is used (fig.\ref{fig:encoding}), quantizing the m/z and intensity values into a unified vector of size \texttt{Dhv} with predefined vectors for m/z (\texttt{ID[0,f]}) and intensity (\texttt{L[0,q]}). Bitwise XOR operations followed by a majority function derive a binarized spectrum \textbf{HV}. Similarity between encoded data points in HDC is measured using Hamming distance, essential for robust classification during inference. We employ data partitioning on the ID and Level arrays with HLS pragmas\cite{amd2024hlspragmas}, enabling concurrent memory accesses.

\subsection{SSD-DRAM Level Data Caching and Optimizations}
\begin{figure*}[!t]
\centering
\includegraphics[width=0.9\textwidth]{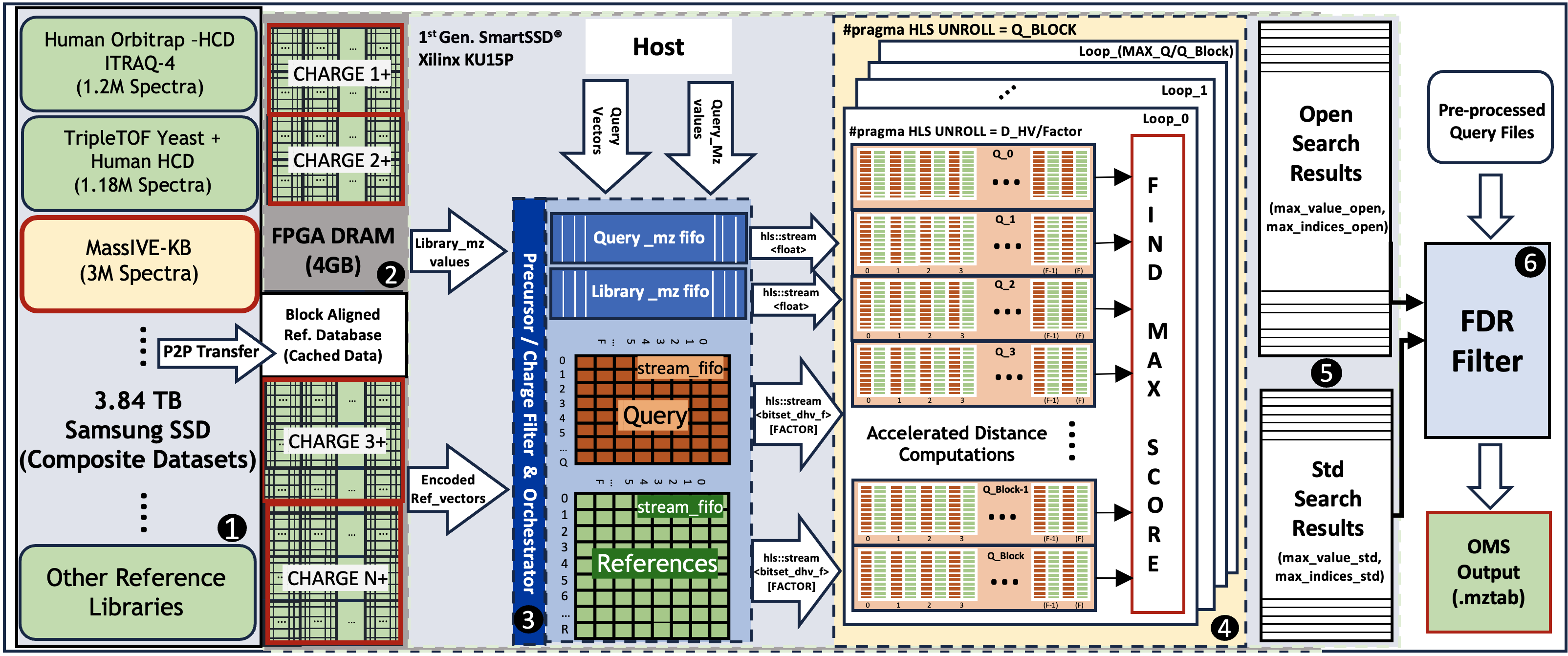} % Adjust the width as needed
\caption{RapidOMS FPGA-Accelerated Spectral Library Searching Flow on Samsung SmartSSD®\vspace{-0.2cm}}
\label{fig:searching_v2}
\end{figure*}

Fig. \ref{fig:searching_v2} provides an overview of the FPGA-accelerated kernel and dataflow of RapidOMS in SmartSSD devices. Once encoded, reference spectra HVs representing large-scale datasets remain static and are processed only once. Given the vast size of MS datasets, each ranging from gigabytes to several hundred gigabytes, it is beneficial to maintain large-scale storage tightly integrated with compute. NAS systems, which are limited by Ethernet bandwidth, lead to inefficiencies in data transfer. In contrast, the Samsung SmartSSD®'s P2P communication\cite{Xilinx2023} feature allows efficient data transfer between NVMe SSD and FPGA, bypassing host memory bottlenecks common in GPU-based tools and significantly reducing latency. As with NAS, SmartSSD systems may also involve stacking storage devices along with processing capabilities. This novel near-storage approach optimizes memory usage by storing pre-encoded HVs in SSD \circled{1} and caching them into DRAM based on their charge states using P2P transfers, especially when the reference database exceeds DRAM capacity. In DRAM, the reference database of encoded vectors is organized by sorted reference precursor m/z (PMZ) values, arranged in block segments, with each block tailored to a specific charge state and structured in blocks of MAX\_R. Each block is defined by its minimum and maximum PMZ values, incorporating the metadata of both query and reference databases along with the open search threshold, enabling independent and streamlined processing. To enhance parallel processing, Q\_BLOCK is introduced as a parallelism factor indicating the number of queries processed concurrently against the reference. Based on Q\_BLOCK and MAX\_R, the orchestrator efficiently directs the structured blocks within the DRAM for retrieval and assigns them for strided access by the FPGA \circled{3}. Serving as a bridge between DRAM storage and FPGA processing, the orchestrator ensures smooth data flow and enhances data retrieval efficiency. Adjusting the threshold variability, guided by the orchestrator, balances search accuracy with efficiency (fig.\ref{fig:effective}).

\subsection{RapidOMS FPGA-accelerated Search Kernel}

After blockwise alignment in DRAM, data is ready for library searching. The reference dataset is retrieved from SmartSSD DRAM, utilizing FPGA’s on-chip memories for fast access. The system processes up to MAX\_R references, corresponding to the established reference block size. The kernel processes MAX\_Q queries in blocks of Q\_BLOCK. Due to the large size of \textbf{HVs}, maintaining their original HV-width is impractical. Therefore, they are divided into multiple streaming FIFOs with smaller bit widths (Dhv/FACTOR), managing their size and ensuring a smooth dataflow-driven pipeline. In the core computation block, we fully harness the parallel processing capabilities of the FPGA. During the initial iteration, reference \textbf{HVs} are cached in the FPGA's on-chip Ultra RAM (URAM) to facilitate frequent comparisons with queries. To optimize performance, we strategically manage Q\_BLOCK for parallel processing of multiple queries \circled{4}. This directly influences the FPGA's resource utilization: higher Q\_BLOCK lead to faster processing (Fig. \ref{fig:qfactor}) but requires a proportionately higher allocation of FPGA resources, needing careful calibration to balance speed and resource availability. Each query, cached in Q\_BLOCK per iteration, operates over a sliding window spanning all the references, processed over multiple loop cycles. At the heart of this process is a fast unrolled XOR and a population count (popcount) module, both parameterized for Dhv/FACTOR running concurrently, showcasing the FPGA's parallel execution capacity. Array partitioning and binding pragmas optimize memory access and balance workload distribution.

The parallelized find\_max\_score function \circled{5} compares query PMZ values with library PMZ values for spectral matching, supporting both standard and open search strategies. In standard search, it checks if the normalized m/z difference is within 20 ppm, updating maximum values and indices if exceeded. The system outputs two result sets: one for standard and one for open search, each including the query number, matched reference IDs, and maximum score.

\subsection{FDR Filter}
In mass spectrometry, accurate spectrum identification is critical, with False Discovery Rate (FDR)\cite{Elias2007} being a key metric. The outputs from the previous step are given to the FDR filtering process. FDR enhancement uses a target-decoy \circled{6} approach, identifying and excluding decoy matches. FDR is calculated as the ratio of decoy to target matches, typically set at a stringent 1\% threshold to minimize false positives. Managing FDR optimizes the balance between sensitivity and specificity, resulting in accurate and reliable identified spectral matches.

\section{RESULTS}

\subsection{Experimental setup}

We tested our workloads on the Gen 1 Samsung SmartSSD® Computational Storage Drive, featuring a 3.2-TB SSD connected to a Xilinx KU15P Kintex UltraScale FPGA with 523K LUTs and 1,045K flip-flops. The FPGA kernel was generated using Vitis HLS 2021.2. Additionally, we evaluated the more powerful Gen 2 SmartSSD® \cite{KachareEmergingCSSSD}, featuring the Versal Prime VM1802 with 1.8x the acceleration resources and 4x the on-chip memory compared to Gen 1. This next-generation device has a Maximum Thermal Design Power (TDP) of 40W, up from 25W in Gen 1. This analysis provides insights into  enhanced capabilities and performance benefits of the latest SmartSSD technology. 

We compare our NS design to the SOTA MS search tools: HyperOMS\cite{hyperoms}(GPU), ANN-Solo\cite{annsolo}(CPU), and SpectraST \cite{Ma2014} focusing on metrics like identifications, speed, and energy efficiency.  These SOTA tools are tested on two GPUs : NVIDIA Geforce GTX 1080Ti (default) and the newer GeForce RTX 4090, as well as Intel i7-8700K with 64GB RAM. Data was sourced from a Synology 50TB DiskStation DS1522+ RAID5 NAS, connected via gigabit Ethernet with a link load of 80\%.

The datasets for analysis included the Yeast+Human HCD spectral library with the iPRG2012\cite{iprg} dataset as the query, and the human spectral library with the b1927-HEK293\cite{hek} dataset as the query (see Table  \ref{table:oms_settings}). 
\begin{table}[h!]
\centering
\tiny
\begin{minipage}[t]{.45\linewidth}
\caption{OMS settings}
\begin{tabular}{lcc}
\hline
\textbf{Dataset} & \textbf{iPRG2012\cite{iprg}} & \textbf{HEK293\cite{hek}} \\
\hline
\# query spectra & 16k & 47k \\
\# ref spectra & 1.16M & 3M \\
Bin size & 0.05 & 0.04 \\
Precursor m/z tol & 20ppm/75Da & 5ppm/75Da \\
FDR threshold & 1\% & 1\% \\
\hline
\end{tabular}
\label{table:oms_settings}
\end{minipage}%
\hfill
\begin{minipage}[t]{.45\linewidth}
\caption{Design Outline}
\begin{tabular}{lc|cc}
\hline
\textbf{Parameter} & \textbf{Value} & \textbf{Resource} & \textbf{Utilization} \\
\hline
MAX\_R & 4096 & LUT & 78.18\% \\
MAX\_Q & 2048 & FF & 32.24\% \\
Q\_BLOCK & 16 & BRAM & 5.60\% \\
Dhv & 4096 & URAM & 56.03\% \\
FACTOR (F) & 16 & DSP & 28.63\% \\
\hline
\end{tabular}
\label{table:resource}
\end{minipage}
\end{table}

\begin{figure*}[!t]
    \centering
    \begin{subfigure}[b]{0.215\textwidth}
        \centering
        \includegraphics[width=\linewidth]{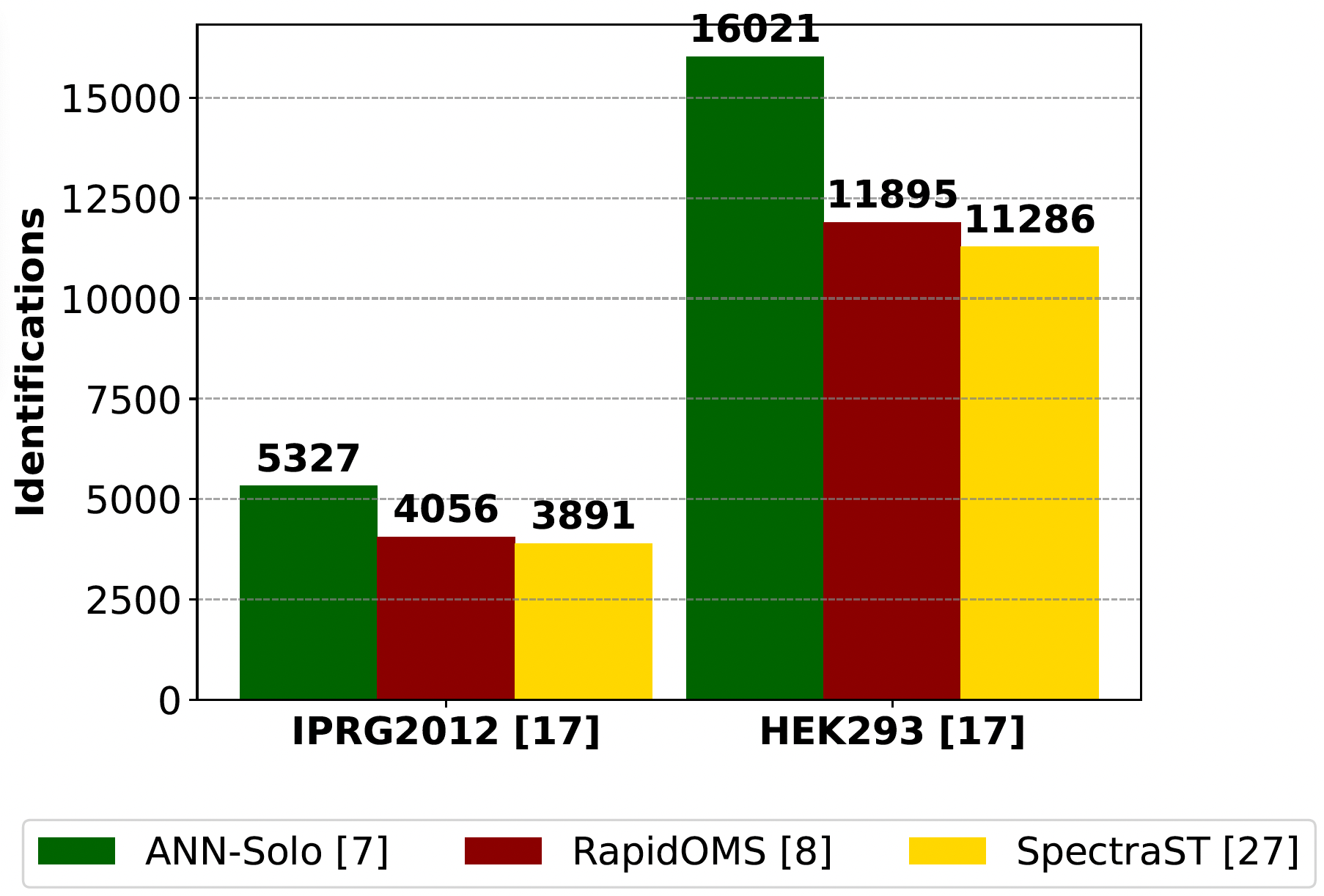}
        \caption{Spectral identifications}
        \label{fig:identifications}
    \end{subfigure}
    \hfill
    \begin{subfigure}[b]{0.18\textwidth}
        \centering
        \includegraphics[width=\linewidth]{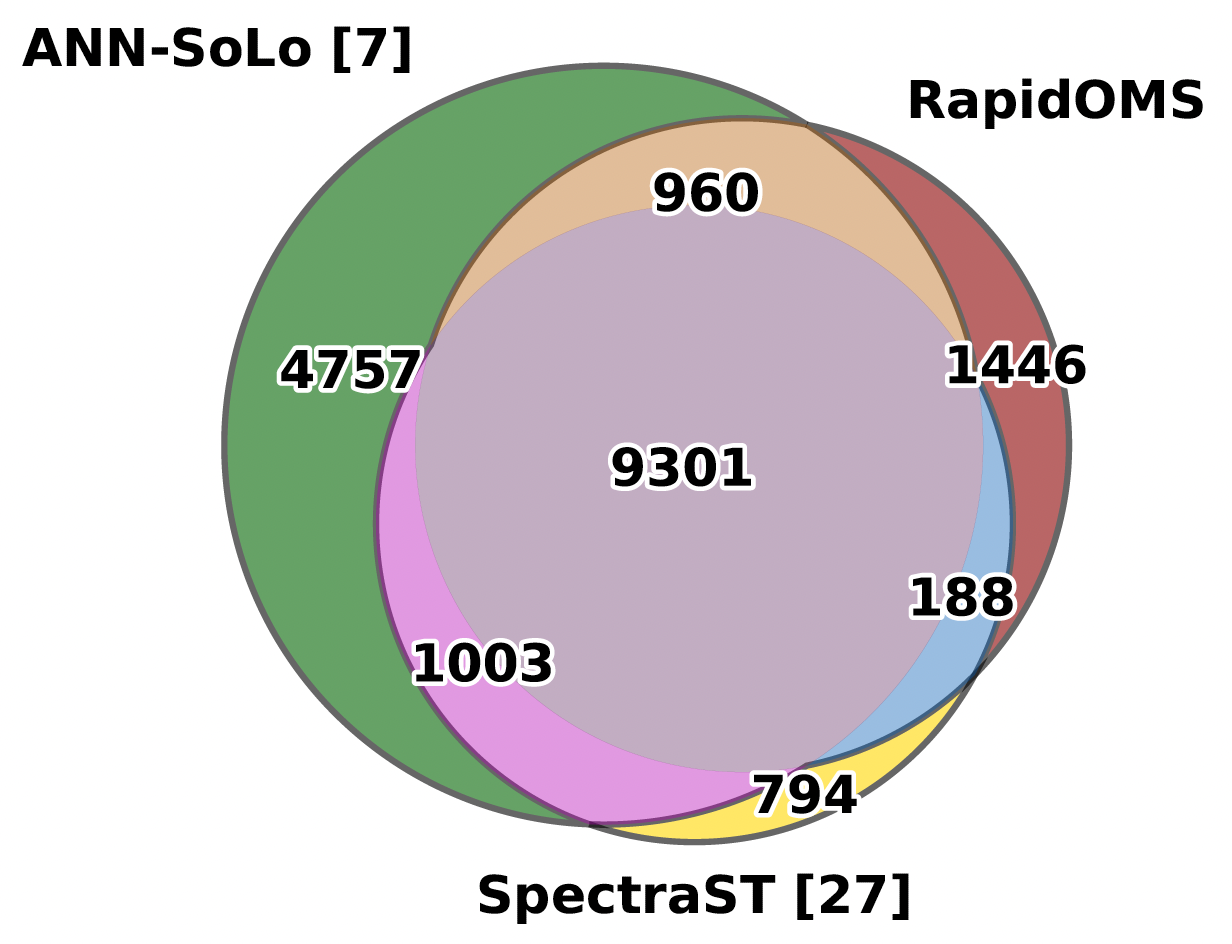}
        \caption{Spectra Consensus}
        \label{fig:1927}
    \end{subfigure}
    \hfill
    \begin{subfigure}[b]{0.20\textwidth}
        \centering
        \includegraphics[width=\linewidth]{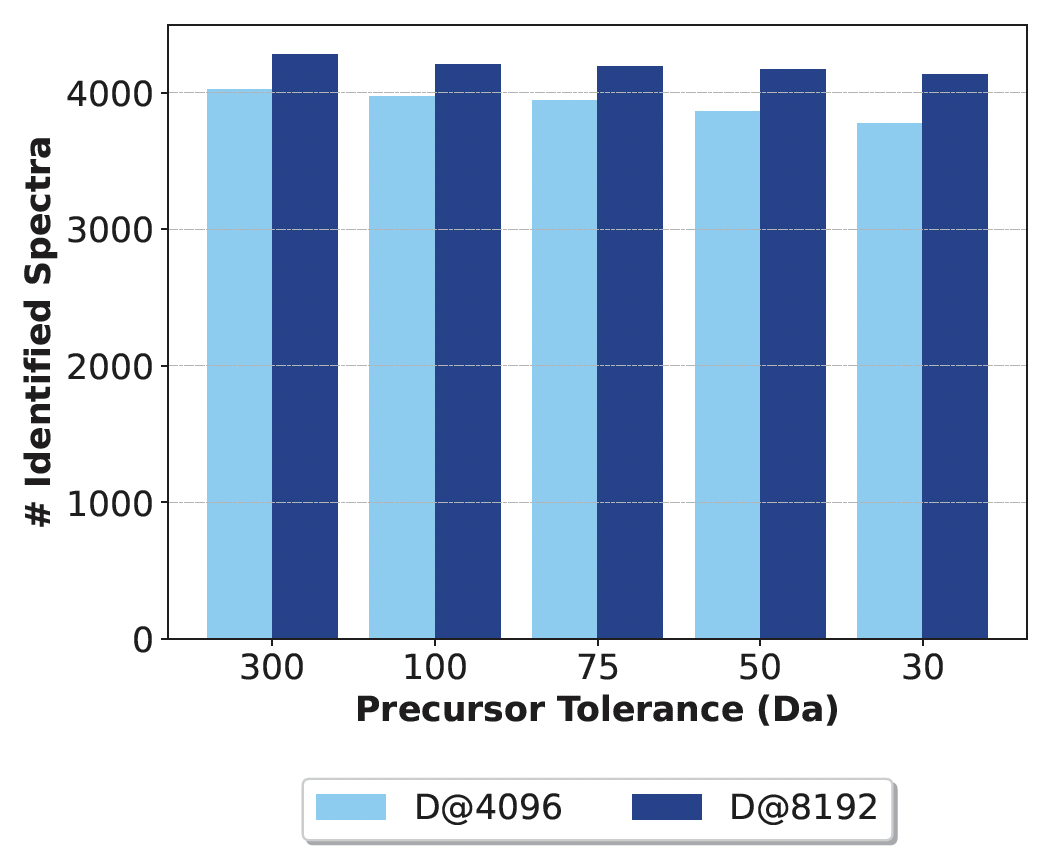}
        \caption{Identification vs PMZ tol}
        \label{fig:dalton}
    \end{subfigure}
    \hfill
    \begin{subfigure}[b]{0.19\textwidth}
        \centering
        \includegraphics[width=\linewidth,height=0.85\textwidth]{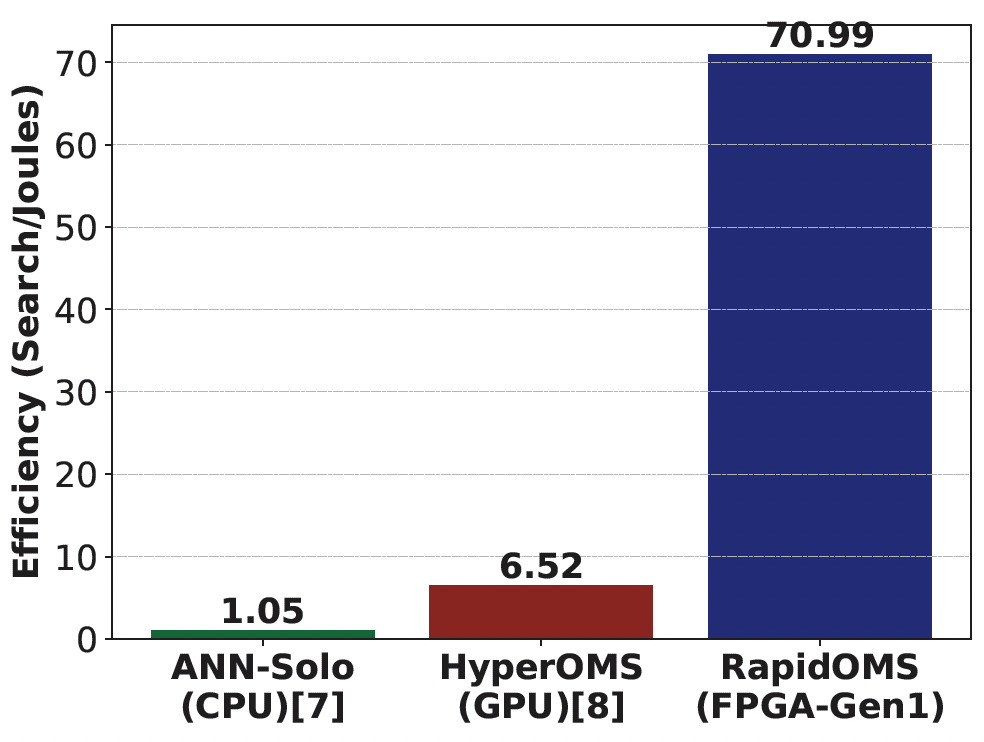}
        \caption{Search Efficiency}
        \label{fig:searchefficiency}
    \end{subfigure}
    \hfill
    \begin{subfigure}[b]{0.19\textwidth}
        \centering
        \includegraphics[width=\linewidth,height=0.85\textwidth]{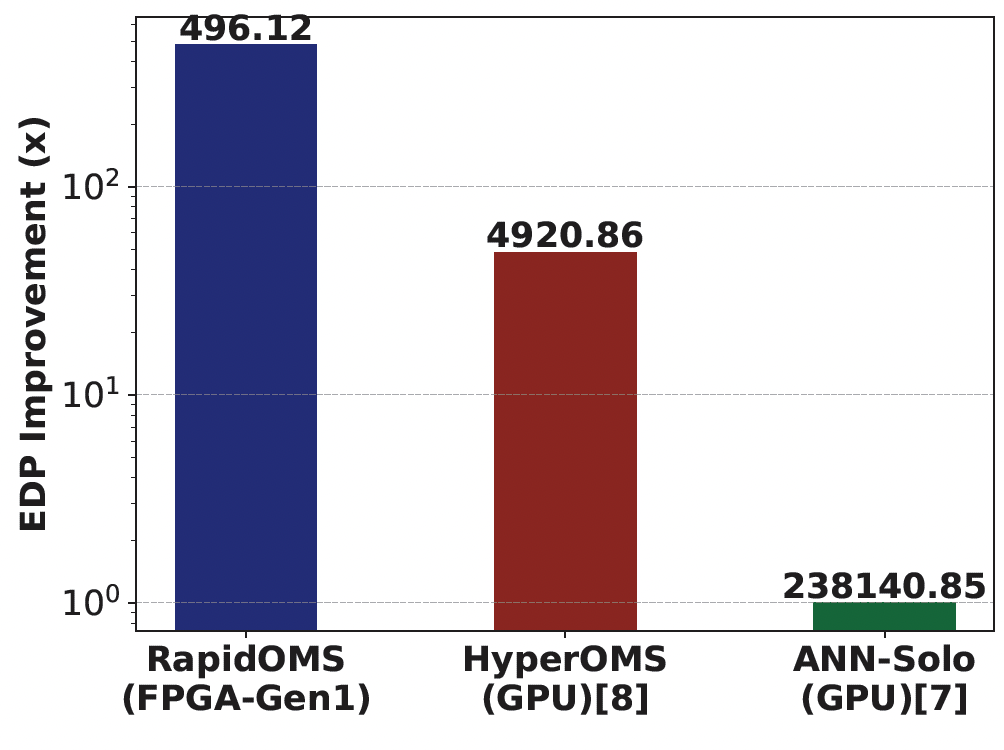}
        \caption{EDP Improvement}
        \label{fig:edpimprovement}
    \end{subfigure}
    \caption{Search Quality and Energy Efficiency\vspace{-0.2cm}}
    \label{fig:search_quality_energy_efficiency}
\end{figure*}

\begin{figure*}[!h]
    \centering
    \begin{subfigure}[b]{0.21\textwidth}
        \centering
        \includegraphics[width=\linewidth]{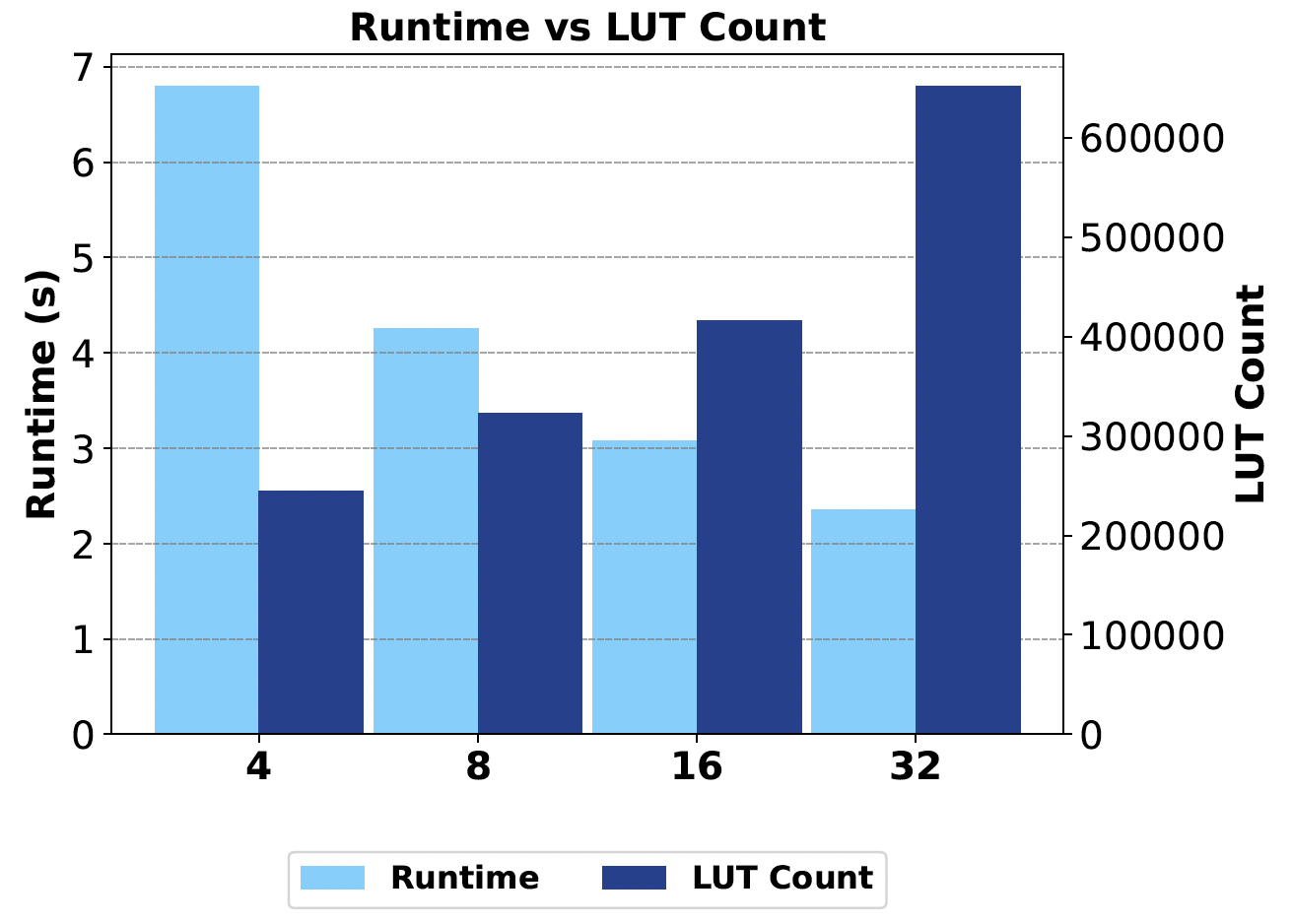}
        \caption{Q-factor Unrolling}
        \label{fig:qfactor}
    \end{subfigure}
    %\hfill
    \begin{subfigure}[b]{0.19\textwidth}
        \centering
        \includegraphics[width=\linewidth]{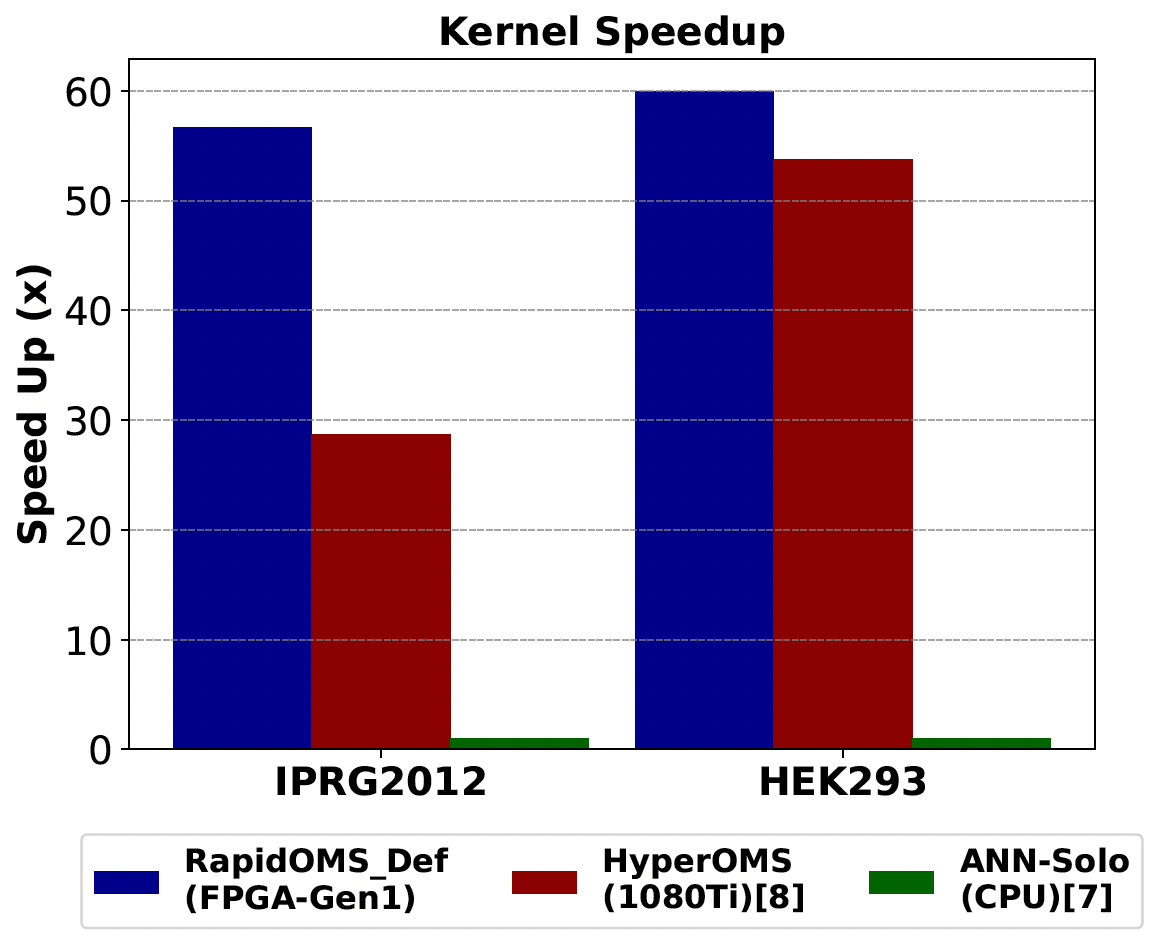}
        \caption{Speedup vs SOTA}
        \label{fig:raw_new}
    \end{subfigure}
    %\hfill
    \begin{subfigure}[b]{0.19\textwidth}
        \centering
        \includegraphics[width=\linewidth]{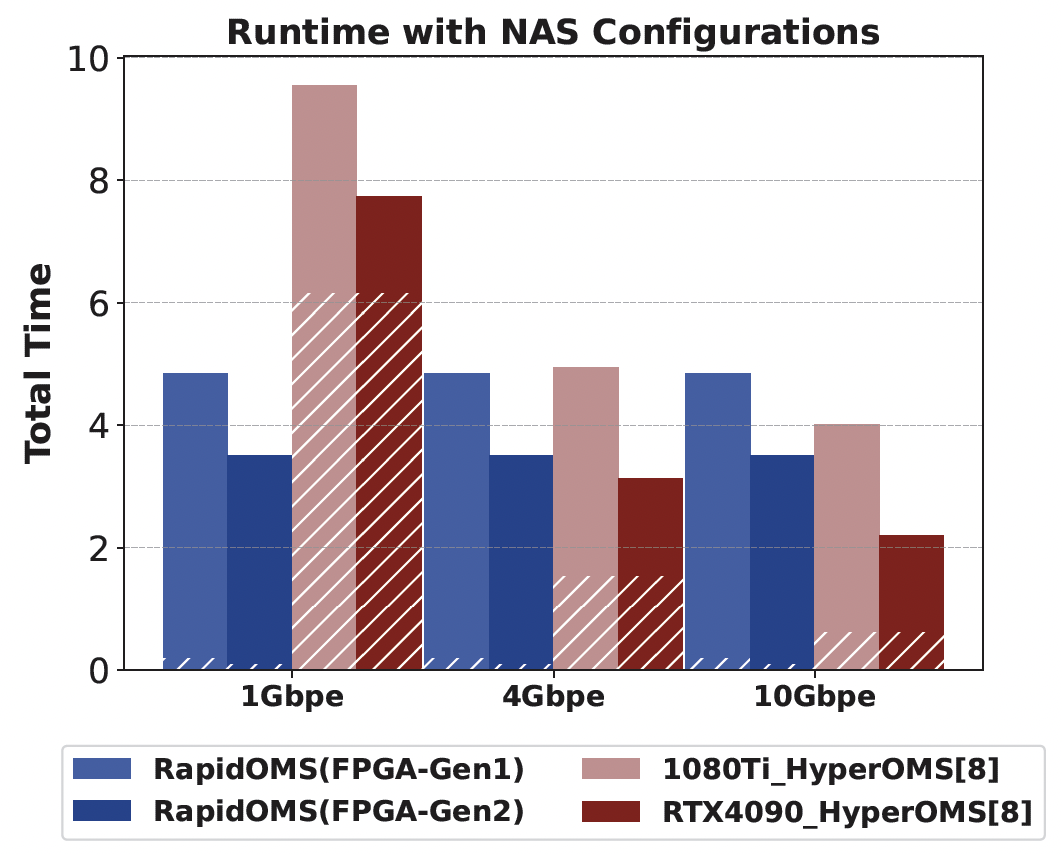}
        \caption{Scaling@ Ethernet BW}
        \label{fig:nas}
    \end{subfigure}
    %\hfill
    \begin{subfigure}[b]{0.195\textwidth}
        \centering
        \includegraphics[width=\linewidth]{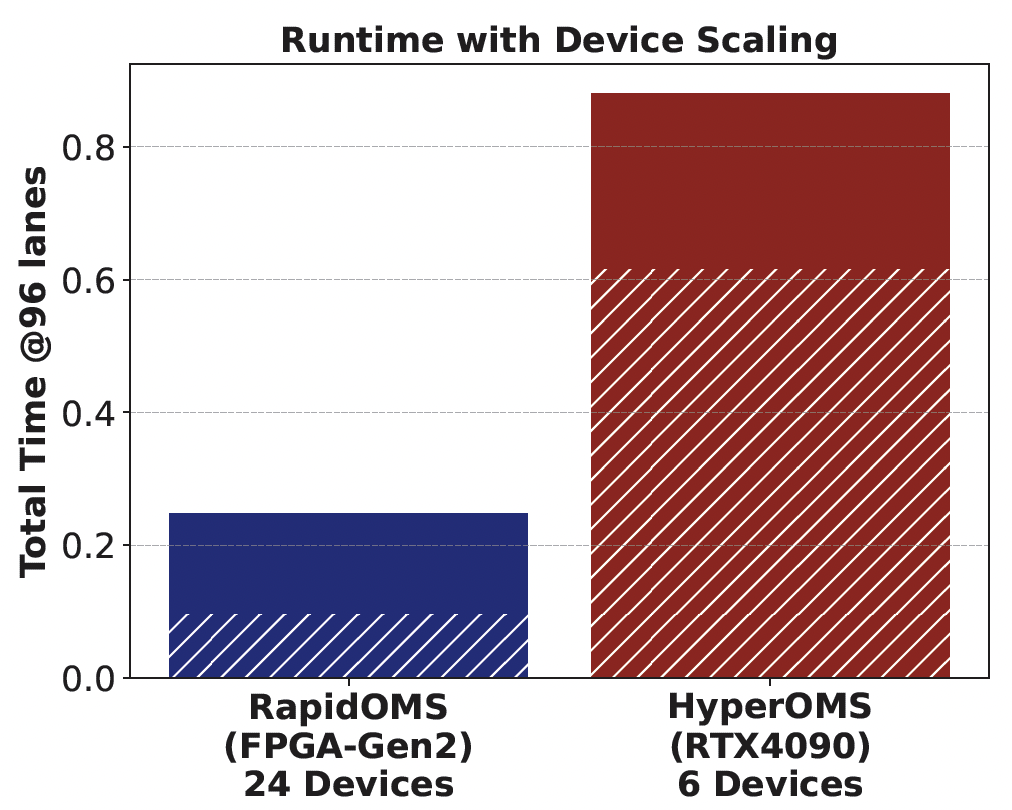}
        \caption{Scaling@ 96 lanes}
        \label{fig:nas_scaled}
    \end{subfigure}
    %\hfill
    \begin{subfigure}[b]{0.19\textwidth}
        \centering
        \includegraphics[width=\linewidth]{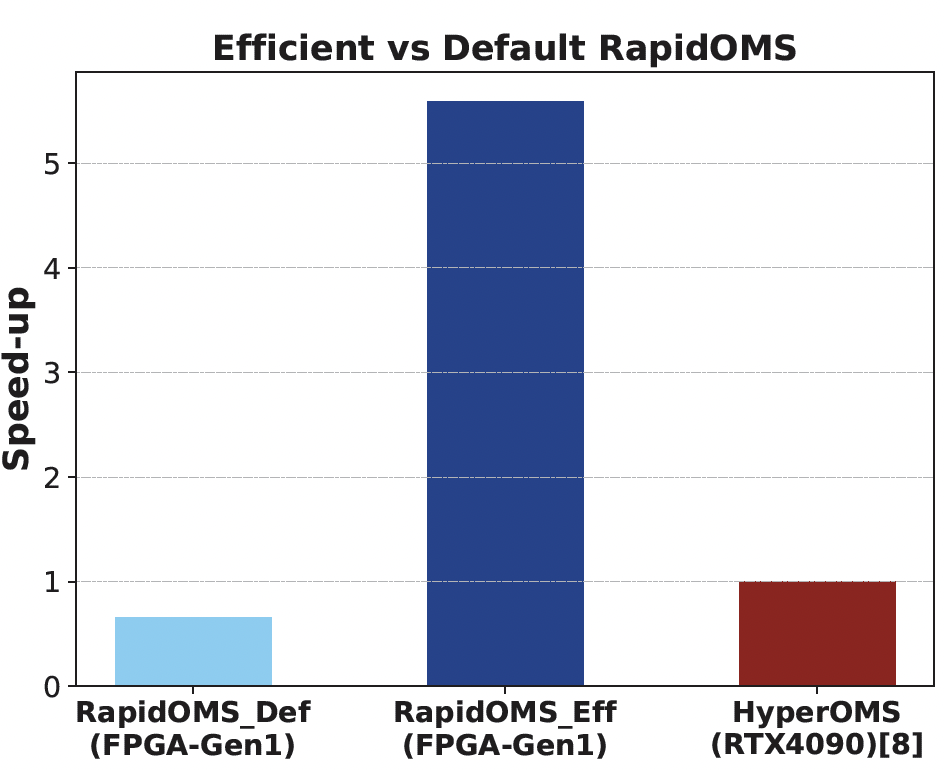}
        \caption{Efficient kernel speedup}
        \label{fig:effective}
    \end{subfigure}
    \caption{Performance and Efficiency Analysis across different configurations \vspace{-0.2cm}}
    \label{fig:performance_efficiency_analysis}
\end{figure*}

\subsection{Search Quality}
We compare search quality of HDC-based tools such as our proposed RapidOMS and HyperOMS\cite{hyperoms} to SOTA tools ANN-SoLo\cite{annsolo} and SpectraST\cite{Ma2014}.  Our benchmark was set against the consensus identifications from various search tools in the iPRG2012\cite{iprg} study. RapidOMS successfully identified 4,056 spectra in the small-scale dataset, surpassing SpectraST’s \cite{Ma2014} 3,891 spectra and compared to ANN-SoLo’s \cite{annsolo} 5,327 identifications. For the more extensive dataset, RapidOMS identified 11,895 spectra, outperforming SpectraST's\cite{Ma2014} 11,286 identifications, yet not reaching the 16,021 spectra identified by ANN-SoLo\cite{annsolo}. But according to Fig. 5(b), RapidOMS not only matched spectra identified by other tools but also uniquely detected spectra that others did not. Previous works indicate identification rates of 33–66\% for human samples\cite{hyperoms}. RapidOMS comfortably maintains a competitive rate within this range, aligning with the current SOTA.

\subsection{Search-Space Efficiency}
Our analysis with the iPRG2012\cite{iprg} dataset (+2 charge) shows that reducing the PMZ dalton (Da) value for the open search parameter significantly improves search-space efficiency, especially at lower Da values, with minimal impact on identification rates (Fig. \ref{fig:dalton}). Based on the design parameters in Table \ref{table:resource}, we selected 75 Da for RapidOMS\_eff's open search range, resulting in a 5.5x kernel runtime speedup compared to HyperOMS and an overall speedup of 9.26x. RapidOMS's ability to tune parameters allows users to balance search quality and efficiency, tailoring the search engine to their specific needs.

\subsection{Performance Improvement Analysis}
RapidOMS on SmartSSD demonstrates significant speedup over the SOTA.  Results up to a Q\_FACTOR of 16 are based on the actual deployment of the iPRG2012\cite{iprg}(charge 2+) dataset on SmartSSD Gen 1, with emulations for Gen 2 SmartSSD handling a Q\_FACTOR of 32. We compared RapidOMS with ANN-Solo and HyperOMS (Fig. (Fig. \label{fig:raw_new}), finding that RapidOMS achieves a remarkable 56x and 60x speedup over the CPU baseline for the iPRG2012\cite{iprg} and b1927-HEK293\cite{hek} datasets, respectively. Although HyperOMS kernel runtime appears 50\% faster due to its nearly 1.6 GHz frequency compared to the 220 MHz achieved by FPGA designs in this specific configuration, accounting for data transfer overheads (highlighted with white stripes) reveals RapidOMS's superior performance—2x faster with iPRG2012\cite{iprg} and 28\% faster with the more query-intensive HEK293\cite{hek} dataset.

Further analysis with the iPRG dataset in Fig.\ref{fig:nas} examines network and storage configurations. We configured the host to receive data over 4GbE and 10GbE ports (@80\% utilization) and assessed NAS read speeds in a RAID5 configuration. At both 4GbE and 10GbE, the RTX4090 surpasses the SmartSSD 2.0 setup, with the 1080Ti trailing, highlighting the impact of network and storage on performance. The advantage of SmartSSD Gen 2 lies in its U2 form factor and PCIe Gen 4 interface, allowing deployment of multiple devices in limited space—key for scalable enterprise solutions. In contrast, GPUs like the RTX 4090 require larger PCIe 4x16 slots. With 96 PCIe lanes, up to 24 SmartSSDs can be accommodated versus only 6 GPUs. This scalability significantly impacts throughput, especially with NAS-to-host data transfers, bottlenecking GPU systems at 1.25 GBps. SmartSSDs benefit from direct NVMe bridge connections, enabling rapid P2P transfers up to 6.4 GBps. 
Even with 10GbE standards, workflows are hindered by data transfer rates, leading to compute inactivity. Fig.\ref{fig:nas_scaled} shows a 4x improvement with RapidOMS’s NS processing, where SmartSSD minimizes data transfer times and enhances efficiency. The FPGA's proximity to storage eliminates long data paths, enabling parallel computing, replacing multiple NAS disks with equivalent SmartSSD capacity.

\subsection{Energy Efficiency}
We observe a stark contrast in power consumption: SmartSSD's modest 23W versus the GPU's substantial 238W, as measured by our profiling tools, Vitis Analyzer and nvidia-smi. Our analysis utilizes the metric of search comparisons per joule to gauge efficiency. RapidOMS on the SmartSSD platform exhibits a significant increase in energy efficiency without compromising the accuracy and integrity of MS analysis, outperforming ANN-Solo\cite{annsolo} by about 68x, primarily due to the CPU's longer runtime. Compared to HyperOMS\cite{hyperoms} on the GPU, RapidOMS demonstrates an 11x improvement in energy efficiency. In HPC environments, where energy and latency are crucial, the Energy Delay Product (EDP) provides a complete measure of energy efficiency. RapidOMS stands out, offering an EDP improvement that is over 480x better than ANN-Solo and 48x better than HyperOMS, underlining its exceptional efficiency and suitability for these critical HPC scenarios.

\section{CONCLUSION}
This paper presents RapidOMS, a novel framework using SmartSSDs that integrates FPGA technology with SSDs for large-scale library searching in MS-based proteomics. It achieves a 60x speedup over CPU baselines and 11x greater energy efficiency compared to GPUs while maintaining a competitive identification rate with current SOTA methods. This balance of speed, efficiency, and accuracy makes RapidOMS a powerful tool for large-scale proteomics analysis with transformative potential for personalized healthcare.

\section*{ACKNOWLEDGMENT}
This work was supported in part by PRISM and CoCoSys, centers in JUMP 2.0, an SRC program sponsored by DARPA (SRC grant number - 2023-JU-3135). This work was also supported by NSF grants \#2003279, \#1911095, \#2112167, \#2052809, \#2112665, \#2120019, \#2211386.

\clearpage 
\bibliographystyle{IEEEtran}
\bibliography{RapidOMS}

\end{document}